\documentclass[twocolumn]{aastex631}

\usepackage{comment}

\newcommand{\barolo}{\textsc{3DBarolo}}
\newcommand{\galpynamics}{\textsc{Galpynamics}}

\newcommand{\kms}{\textrm{km s}^{-1}}
\newcommand{\hi}{\text{H\,\sc{i}}}
\newcommand{\Msol}{\textrm{M}_{\odot}}
\newcommand{\wkapp}{\textsc{WKAPP}}
\newcommand{\sofia}{\textsc{SoFiA-2}}

\newcommand{\ND}[1]{\textcolor{blue}{#1}}

\newcommand{\R}[1]{\textcolor{red}{#1}}

\newcommand{\queens}{Department of Physics, Engineering Physics, and Astronomy,Queen's University, Kingston ON K7L~3N6, Canada}
\newcommand{\ICRAR}{International Centre for Radio Astronomy Research (ICRAR), The University of Western Australia,\\ 35 Stirling Highway, Crawley WA 6009, Australia}

\newcommand{\CSIRO}{CSIRO Space and Astronomy, PO Box 1130, Bentley WA 6102, Australia}

\defcitealias{Deg2024}{D24}
\defcitealias{ManceraPina2019}{MP19}

\begin{document}

\title{A diffuse dwarf on the bTFR}
\title{\textcolor{black}{WALLABY Pilot Survey: A gas-rich diffuse dwarf on the baryonic Tully Fisher relation}}

\author[0009-0003-3774-3430]{Rebecca Dudley}
\affiliation{David A. Dunlap Department of Astronomy, University of Toronto, Toronto ON M5S 3H4, Canada}
\affiliation{\queens}

\author[0000-0003-3523-7633]{N. Deg}
\affiliation{\queens}

\author{Kristine Spekkens}
\affiliation{\queens}

\author[0000-0002-3929-9316]{N. Arora}
\affiliation{Arthur B. McDonald Canadian Astroparticle Physics Research Institute, Queen's University, Kingston, ON K7L 3N6, Canada}
\affiliation{\queens}

\author{T. O'Beirne}
\affiliation{Centre for Astrophysics and Supercomputing, Swinburne University of Technology, Hawthorn, Victoria 3122, Australia}
\affiliation{European Southern Observatory, Karl-Schwarzschield-Str. 2, 85748 Garching near Munich, Germany}
\affiliation{\CSIRO}

\author[0000-0003-3636-4474]{V. Kilborn}
\affiliation{Centre for Astrophysics and Supercomputing, Swinburne University of Technology, Hawthorn, Victoria 3122, Australia}

\author[0000-0002-7625-562X]{B. Catinella}
\affiliation{\ICRAR}

\author[0000-0001-5175-939X]{Pavel E. Mancera Piña}
\affiliation{Leiden Observatory, Leiden University, P.O. Box 9513, 2300 RA, Leiden, The Netherlands}

\begin{abstract}
Diffuse dwarf galaxies, and particularly ultra diffuse galaxies (UDGs), challenge our understanding of galaxy formation and the role of dark matter due to their large sizes, low surface brightness, and varying dark matter content. In this work, we investigate the gas-rich diffuse dwarf galaxy WALLABY J125956-192430 (aka. KK176) using high-resolution \hi\ data from the WALLABY survey. 
We produce the most reliable kinematic model for KK176 to date. Using this model, the derived mass decomposition shows that KK176 is dark matter dominated. We also place KK176 on the baryonic Tully-Fisher relation (bTFR), finding that it is consistent with low-mass dwarf galaxies but distinctly different from reported dark matter-deficient UDGs. 
\end{abstract}


\section{Introduction} \label{sec:intro}

The properties of dwarf galaxies have the potential to constrain models of cosmological galaxy formation and the role of dark matter (DM, e.g. \citealt{Sales2022}).
The structure of diffuse, low-surface brightness star-forming  systems are well-known to challenge these models (e.g. \citealt{KuzioDeNaray2006,KuzioDeNaray2011}).
The most extreme such systems are field ultra-diffuse galaxies (UDGs). While they were originally defined as having
$\mu_{g,0} \geq 24,$mag arcsec$^{-2}$ and $R_e \geq 1.5,$kpc \citep{van_dokkum_2015}, this definition is sensitive to the limiting angular resolution of surveys such as Dragonfly. More recent work has introduced definitions based on scaled relations (e.g. \citealt{Lim2020}, from the NGVS survey).
Unlike some classes of similar objects in denser environments \citep{Buzzo2025}, an increasing amount of evidence suggests that gas-rich, star forming UDGs and diffuse dwarfs have similar properties (e.g. \citealt{Jones2023,Motiwala2025,Wright2025}.

The minor stellar contribution to the baryonic contents of diffuse dwarfs (and therefore smaller uncertainties due to stellar mass-to-light ratios) has the potential to provide a clearer view of their DM dynamics than in higher mass systems. So far, however, an inconsistent picture has emerged: some reports of DM-deficient field objects \citep{mancera_pina_2022, Sengupta2019}
are hard to reconcile with cosmological galaxy formation, while others \citep{Scott2021} find DM halos that are proportional to other systems of similar masses, which is in agreement with cosmological gaalxy formation.  
The DM contents of UDG and diffuse dwarf halos, and how those
halos compare with those in the broader galaxy population, is a crucial piece of the puzzle to connect them to cosmology and decipher their structure and evolution.

The \hi\ disks of gas-rich galaxies have long been used to constrain DM halo structure through
mass models of their rotation curves \citep{Oh2015, Read2016, mancerapina_2025}. However, very few diffuse dwarfs currently have \hi\ maps that are detailed enough to conduct the same studies: at least four resolution elements (beams) across the disk major axis are needed to separate disk geometry from rotation via standard 3D tilted-ring (TR) modeling techniques \citep{Deg_2022}, a threshold that is not often met for diffuse dwarfs and UDGs. However, for some cases, such as \citet{Sengupta2019, ManceraPina2019} (hereafter, \citetalias{ManceraPina2019}), there are sufficient elements. A further challenge is that the priors on the disk geometry from optical estimates are notoriously unreliable for low-mass galaxies, particularly at the relatively low inclinations towards which diffuse dwarf searches are biased \citep{Read2016,Banik_2022}. Reliable kinematic models that constrain both disk rotation and geometry are therefore crucial for measuring DM properties in diffuse dwarfs.

This challenge is shown in recent efforts to place gas-rich diffuse dwarfs on the Baryonic Tully-Fisher
Relation (bTFR), a fundamental scaling relation between the circular velocities and baryonic masses of galaxies that investigates the interplay between DM and
baryons inside galaxies. Whether or not UDGs lie on the bTFR is not yet clear, with some studies arguing for systematic deviations implying low concentration DM halos (\citetalias{ManceraPina2019}; \citealt{Hu2023,Du2024}), while others report consistency \citep{He2019,Karunakaran2022}. A similar discussion is underway for low-mass dwarf galaxies, for which evidence of an upturn at the low-velocity end of the bTFR has been claimed for some samples \citep{McQuinn2022} but not others \citep{Giovanelli2013, Iorio2017, mancerapina_2025}. The need to reliably place additional diffuse dwarfs on the bTFR to address these discrepancies is therefore clear.


In this context, LEDA~44681, detected in \hi\ as WALLABY~J125956-192430 in the untargetted Widefield ASKAP L-band Legacy All-sky Blind surveY (WALLABY) \citep{Koribalski2020,Hotan2021,Murugeshan2024} and named KK176 in the  discovery, is particularly interesting to examine. Previously studied as a nearby, extremely metal poor void galaxy \citep{Pustilnik2020,Pustilnik2021,Kurapati2024}, KK 176 has a primary distance measurement of $d = 7.3 \pm 0.3\,$Mpc from the tip of its Red Giant branch \citep{Karachentsev17} and is in a void, making it one of the nearest isolated gas-rich diffuse dwarfs.

The WALLABY \hi\ detection of KK 176 is sufficiently spatially resolved to be kinematically modeled by the automated pipeline described by \citet{Deg_2022}. This suggested an \hi\ morphology and kinematics consistent with a high-inclination ($i \sim 60^\circ$) rotating disk, which \citet[][\citetalias{Deg2024}]{Deg2024} used to estimate \hi\ structural parameters using an approximate flow model distance \citep{Kourkchi2020} for consistency across the sample. The proximity, well-constrained distance, relatively high disk inclination, and gas-rich, diffuse nature of KK 176 make it an ideal system to investigate further in order to estimate its DM content and place it on the bTFR.

In this paper, we improve upon the \citetalias{Deg2024} structural models for KK 176 by combining WALLABY \hi\ data with other multi-wavelength datasets.  The core goal of this work is measuring the DM content of KK 176 and placing it on the bTFR. Sec. \ref{sec:data} presents the data that we use for our analysis.  Sec. \ref{sec:modelling} describes our procedure for modeling the \hi\ kinematics and the stellar morphology of KK 176.  Sec. \ref{sec:DMcontent} then uses this model to calculate the DM content of KK 176 and place it on the bTFR, along with other dwarf galaxies in the literature, and Sec. \ref{sec:conclusion} presents our conclusions.

\section{Data} \label{sec:data}

The WALLABY \hi\ observations of KK 176 have an angular and spectral resolution of $30 \arcsec$ and $18.5~\rm{kHz}$ (=$3.9~\kms$ at the observation centre) respectively, with a noise level of $2.6~\textrm{mJy\, beam}^{-1}$.  
This object was detected in the WALLABY PDR2 observations \citep{Murugeshan2024} of the NGC 5044 field using the \hi\ Source Finding Application (\sofia; \citealt{Serra2015,Westmeier2021}), which generated the cubelet and mask used for the rest of this analysis.  Table \ref{Fluffy_Table} lists both the core properties of KK 176 as well as quantities derived in this work (see Secs. \ref{sec:modelling}-\ref{sec:DMcontent}).

\begin{figure*}
    \centering
    \includegraphics[width=0.8\linewidth]{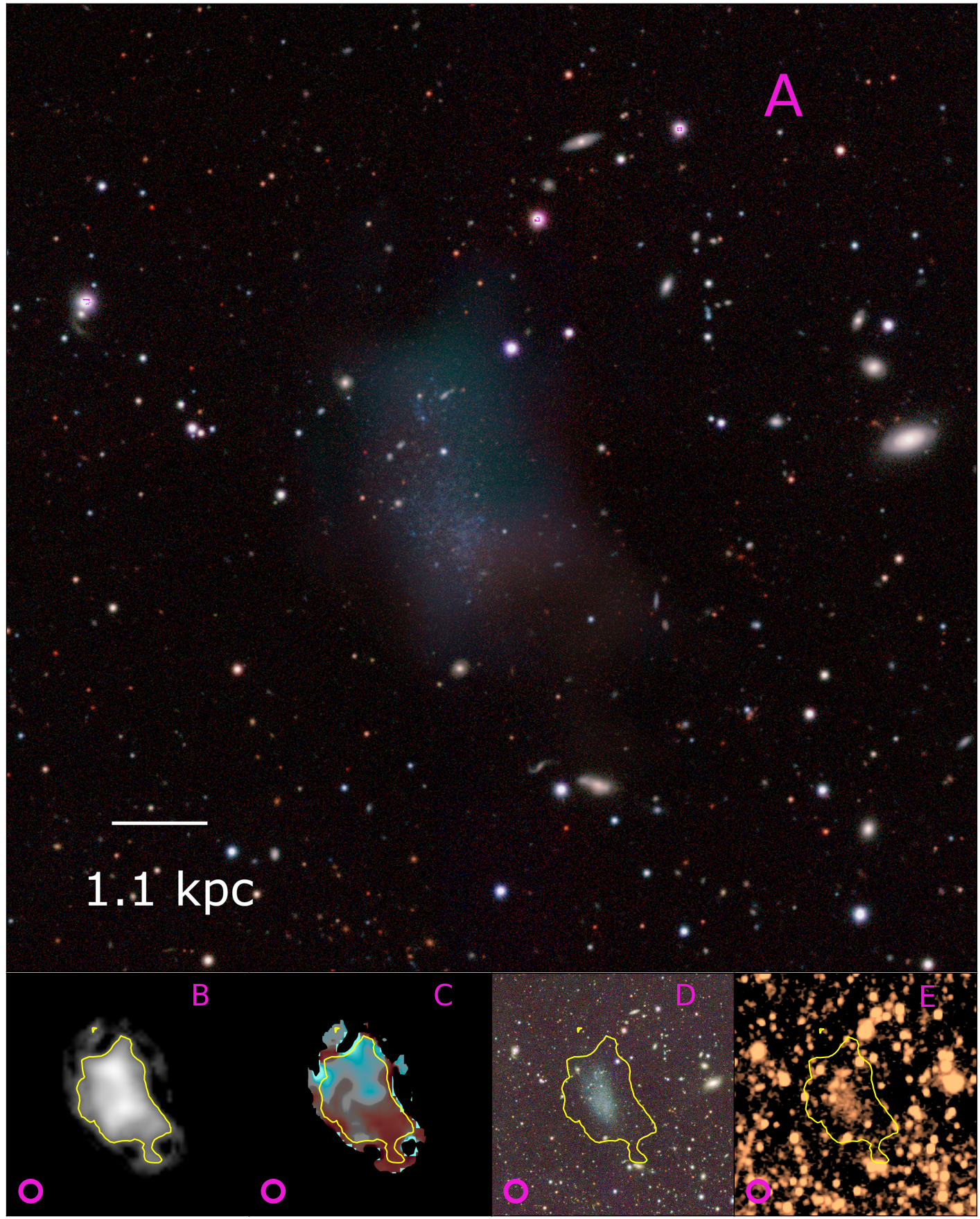}
    \caption{KK 176 as seen at various wavelengths.  Panel A shows the \hi\ component overlayed on a $grz$ composite optical image from DECaLS. The \hi\ color is mapped to the velocity (blue approaching, red receding relative to the estimated $V_{\rm{sys}}$) with the brightness mapped to the intensity.  Panels B and C show the integrated WALLABY \hi\ intensity and line-of-sight velocity maps, respectively.  Panel D shows the optical image, using a different stretch from Panel A to highlight low surface brightness features.  Panel E shows the WISE W1 image.  In Panels B-E the magenta circle shows the WALLABY beam (which is the same as the scale line in Panel A) and the yellow contour shows the  $\Sigma_{\rm{HI}}=1~\Msol~\rm{pc^{-2}}$ \hi\ surface density contour level of the integrated intensity map.
    }
    \label{fig:overlay}
\end{figure*}

In addition to the WALLABY detections, we utilize $g$, $r$, and $z$ band cutouts from the Dark Energy Camera Legacy Survey (DECaLS, \citealt{DESI,Dey2019}) and W1 band cutouts from the Wide-field Infrared Survey Explorer (WISE; \citealt{Wright2010}) to study the stellar content of KK 176.  The WISE cutouts were also obtained using the DECaLS skyviewer. Figure \ref{fig:overlay} shows an overlay of the \hi\ gas onto a 3 colour image generated from the DECaLS cutouts\footnote{CARTA, \citep{CARTA}, and GIMP, \\ \citep{GIMP}.} (Panel A) as well as individual moment maps and images (Panels B-E).


There are several key things to note from this overlay.  Firstly, panels A and D of Figure \ref{fig:overlay} show that the \hi\ extends well beyond the optical extent of KK 176. Additionally, the $1~\Msol~\rm{pc}^{-2}$ contour (in yellow) highlights that the optical is not precisely centered on the brighter, blue emission.
This complexity highlights the challenges faced when modelling a low mass diffuse dwarf.

In terms of stellar emission, KK 176 is predominately blue and has a complex, almost triangular appearance on the sky (Panel D of Figure \ref{fig:overlay}).  This is relevant for determining the shape and structure of the galaxy and measuring the stellar surface brightness profile (see Sec. \ref{sec:modelling} for details).  One reason for this complexity is the stochasicity of star formation in low surface brightness galaxies \citep[e.g.][]{Jones2025}.  Localized star formation will appear blue and dominate the surface brightness profile, making the determination of the galaxy's geometry challenging and biasing standard approaches \citep{Read2016,Banik_2022}.  The WISE photometry, however, shows the older stellar population, providing a more reliable measure of the center point than either the \hi\ or optical observations.  By comparing the outer \hi\ distribution to the W1 band photometry, it is revealed that the \hi\ and WISE centroids are less offset than the \hi\ to optical observations.  Unfortunately, the WISE photometry is complicated by bright foreground sources in the north-east (see Panel E of Figure \ref{fig:overlay}), making it unsuitable for the determination of KK 176's inclination and position angle.  In this case, the most reliable measures of the inclination and position angle are from kinematic models of the \hi\ distribution. 

In addition to this data, \citet{Karachentsev17} obtained HST observations of KK 176.  They used these to obtain a tip of the red giant branch (TRGB) distance of $7.3\pm0.3~\rm{Mpc}$.  This is significantly different than the ($9.9~\rm{Mpc}$) distance used in both \citetalias{Deg2024} and Deg et al. (submitted), who determined the distance using Cosmic Flows-3 \citep{Kourkchi2020}. The larger distance used in those works means that their measures of the \hi\ and baryonic masses will be significant overestimates.  Those works calculate the \hi\ mass using the corrected \hi\ flux provided in \citet{Murugeshan2024}.  We have instead adopted the total \hi\ flux of $4.49\pm0.14~\rm{Jy}~\kms$ from \citet{Kurapati2024}.  This flux, which is from the Giant Metrewave Radio Telescope (GMRT) observations, does not require a statistical correction.  As such, the total flux measurement and associated uncertainty is more reliable than the total corrected WALLABY flux.
At a distance of $7.3\pm0.3~\rm{Mpc}$, the \citet{Kurapati2024} flux gives a total \hi\ mass of $\log_{10}(M_{\hi}/M_{\odot})=7.75\pm 0.05$.


\begin{table} \label{Fluffy_Table}
\centering
\begin{tabular}{lc}
\hline
\hline
{Parameter}                                     & {Value}                       \\ \hline
Cross-Match ID                                & LEDA 44681                              \\
RA (J2000)                           & 12h59m56.3s \\
Dec (J2000)                          & $-19^{\circ}24'48''$ \\
\textit{d}                   & $7.3 \pm 0.3$ Mpc                                     \\ \hline
$i$                                           & $(58\pm~10)^\circ$                                    \\
PA                                            & $(203~\pm~10)^\circ$                                 \\
$V_{\rm{sys}}$                                     &$ (827 ~\pm~4)~\kms$                                 \\
$R_{\rm{\hi}}$           &  $(1.8\pm 0.2) ~\rm{kpc}$  \\
$V_{\rm{\hi}}$   & $(16 \pm 3)~\kms$  \\
$V_{\rm{circ}}(R_{\hi})$ & $(18 \pm 3)~\kms$  \\
$\mu_{0,g}$                                   & $(24.30 \pm 0.03)$ mag arcsec$^{-2}$                                   \\
$R_e$                                         & $(1.30 \pm 0.08)$ kpc                                    \\
$\log_{10}\left({M_*}/{M_\odot}\right)$    & 6.8 $\pm$ 0.2                          \\
$\log_{10}\left({M_{\hi}}/{M_\odot}\right)$ & 7.7 $\pm$ 0.1                         \\
$\log_{10}\left({M_{\rm{B}}}/{M_\odot}\right)$ & 7.9 $\pm$ 0.1                         \\
\hline
\end{tabular}
\caption{A compilation of the various properties of KK 176. RA and Dec are measured from the WISE data (see Sec. \ref{ssec:KinModel}), and held fixed in our preferred kinematic and stellar models. 
$d$ is the TRGB distance from \citet{Karachentsev17}.  The disk inclination, $i$, position angle, $PA$, systemic velocity, $V_{\rm{sys}}$, size $R_{\rm{\hi}}$, and \hi\ velocity $V_{\rm{HI}}$~=~ $V_{\rm{m}}(R_{\rm{HI}})$ were determined from our kinematic model (see Sec. \ref{ssec:KinModel}).  The total circular speed due to gravity, $V_{\rm{circ}}(R_{\hi})$ is the sum of the \hi\ velocity and the asymmetric drift in quadrature (see Sec. \ref{ssec:massmodel}).  The central surface brightness $\mu_{0,g}$, half-light radius $R_e$, and total stellar mass $M_*$ are determined from our stellar model (see Sec. \ref{subsec:stellar_models}).  We have adopted the total \hi\ mass $M_{\hi}$ from \citet{Kurapati2024}. The total baryonic mass is given by $M_{\rm{Bary}}=M_{*}+1.35 M_{\hi}$, 
where the factor of 1.35 is meant to account for helium and other metals \citep{Arnett1999,Oh2015}. It is worth noting that KK 176 is known to be metal poor \citep{Pustilnik2020,Pustilnik2021,Kurapati2024}, which suggests that any molecular gas component will likely be low mass.
}
\end{table}

\section{Modelling} \label{sec:modelling}

Modelling the DM content of a galaxy requires both a kinematic model as well as a stellar density model.  In this section we describe our modelling of KK 176.  We also compare our new kinematic model to the existing model from the WALLABY pilot data release 2 (see \citealt{Murugeshan2024} and \citetalias{Deg2024}).  The previous model was based on only the \hi\ content and did not consider the stellar component of the galaxy; the new model benefits from a better center based on the WISE images.  Moreover, the model was derived using the Cosmic-Flows 3 distance rather than the TRGB distance we have adopted.  Our goal is to build a fully self-consistent model across both the \hi\ and stellar components, which is not the case for the KK 176 model of \citetalias{Deg2024}.


\subsection{Kinematic Model} \label{ssec:KinModel}

There are many kinematic modelling software packages available \citep{Spekkens2007,Jorza2007,Kamphuis2015,diTeodoro15,Deg_2022} for \hi\ observations.  We have chosen to utilize the 3D-Based Analysis of Rotating Objects From Line Observations (\barolo\footnote{\url{https://editeodoro.github.io/Bbarolo/}}; \citealt{diTeodoro15}) code for this analysis.  \barolo\ is a 3D TR modelling code that treats a galaxy as a series of nested rings.  In a TR model, the line of sight velocity at any given point in a ring is given by 
\begin{equation}
V_{\rm{los}}(R,\theta) = V_{\rm{sys}} + V_{\rm{m}}(R)\sin i \cos \theta~,
\end{equation}
where $V_{\rm{sys}}$ is the systemic velocity, $V_{\rm{m}}$ is the rotational velocity of the model's ring, $\theta$ is the position angle of the ring relative to the major axis, and $i$ is the inclination angle (from 0$^\circ$ to 90$^\circ$, where 90$^\circ$ is an edge-on-view).  A particular strength of \barolo\ is the ability to both supply initial estimates or fully fix model parameters, including the centroid, inclination, and position angle.  This means that the center point can be determined using the WISE observations.

The first step in our modelling process is the determination of a common stellar and gaseous center point.  To do this, we applied the \textsc{AutoProf} software package \citep{Stone2021} to the WISE W1 image. This ensures that the center point is matched to the underlying old stellar population. \textsc{AutoProf} performs a non-parametric fitting of both the stellar surface brightness profile and the center point. The full details of the centering can be found in \citet{Stone2021}.  In brief, \textsc{AutoProf} samples a set of apertures based on the PSF, and then refines the initial estimate via a Nelder-Mead simplex optimizer to find the center point to below a single pixel.
Taking the average direction, flux values are sampled along a line from the centre out to 10 times the PSF. 

To generate KK 176's kinematic model, we used the WALLABY data, fixed the \hi\ center to the WISE center and used \barolo\ to fit the \hi\ distribution.  We only considered `flat' disk models where the inclination and position angle are constant across all radii. It is known that measuring the inclination for low resolution and low $S/N$ observations is challenging (see for example \citealt{Deg_2022,Mancera_Pina_2024}.  However, Deg et al. (submitted) explored the reliability of low resolution, low $S/N$,  low rotation kinematic models and found that, \barolo\ models with measured inclinations $\ge40^{\circ}$ tend to be reliable.  In order to test this further, we generated \barolo\ models of different initial inclination guesses, however this did not have dramatic effects on the resultant model rotation curves. We also fixed the velocity dispersion to $V_{\rm{disp}}=8~\kms$.  \citet{Deg_2022} demonstrated that velocity dispersion is poorly constrained for WALLABY observations and $8~\kms$ is the lower value adopted by \citet{Murugeshan2024}. Gas-rich dwarfs in the mass range of KK176 also show typical values of $V_{\rm{disp}}$ around 8 km/s \citep{Iorio2017, mancerapina_2025}. We tested a variety of differing velocity dispersions and found no significant variations in the resulting models.  

\begin{figure}
    \centering
    \includegraphics[width=1\linewidth]{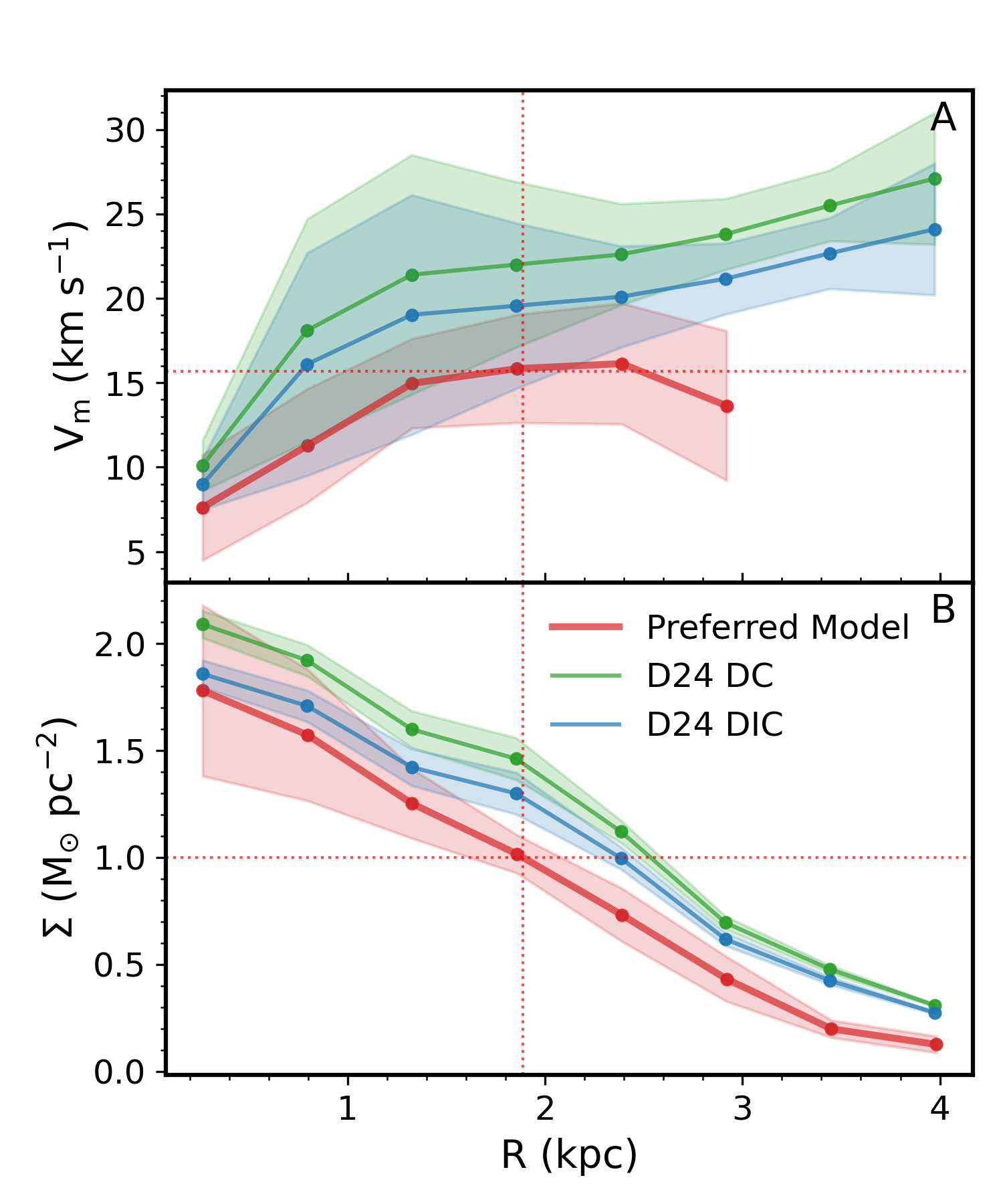}
    \caption{Our preferred kinematic model of KK 176 (red curves), in comparison to the pipeline-generated model presented by \citetalias{Deg2024} (blue and green curves). Shaded regions represent the uncertainties on the curve of the corresponding color. Panel A shows rotation curves, and panel B shows surface density profiles.
    The horizontal dotted line in panel B shows an \hi\ surface density of $\Sigma_{\rm{HI}}=1~\Msol/\rm{pc}^2$, from which we define $R_{\rm{HI}}$, represented by the vertical dotted line line (see Sec.~\ref{sec:modelling}). The horizontal dotted line in panel A shows $V_{\rm{HI}}$. The green curve (= DC) is the D24 model corrected to KK 176's TRGB distance $d$. The blue curve (=DIC) is the same initial model when corrected for both $d$ and $i$ from our preferred model (see Table~\ref{Fluffy_Table}).}
    \label{fig:RC_and_SD}
\end{figure}

Figure \ref{fig:RC_and_SD} shows the rotation curve for our preferred \barolo\ model and the surface brightness profile extracted from the data (in red), and Table \ref{Fluffy_Table} lists the best fitting geometric parameters. For clarity, we follow \citet{Deg_2022}, and set the radial bins in \barolo\ to be $15\arcsec$ in size (i.e 0.5 beams).  It is worth noting that the different radial extents of our preferred model and the distance corrected \citetalias{Deg2024} model is solely due to the difference in the center point.  However, the differing models have similar extents for their surface density profiles due to the fact that \barolo\ calculates the surface density profiles via ellipses on the moment 0 map.
To check our fit, Appendix \ref{app:ChannelMaps} shows a comparison of the model channel maps to the underlying data.  These maps show that the model and data have residuals expected for an object with as low of a $S/N$ as KK 176.  They do not show any systematic differences that would point to a model failure.

In order to robustly measure the DM content of KK 176 it is necessary to propagate uncertainties in the underlying model parameters to the DM measures.  For this reason, we have adopted the `flipping' bootstrap technique described in Appendix \ref{apdix:bootstraps}.  We used this to generate 100 independent bootstrap samples and then fit each of those samples using \barolo.  The distribution of these fits provides the uncertainties for KK 176 in Figure \ref{fig:RC_and_SD}. 

From a given kinematic model, it is possible to extract the model velocity and size parameters, while the \hi\ surface density profile is taken directly from the data and corrected for the chosen inclination rather than being modeled.
 The size, $R_{\hi}$, is the radius where surface density is $\Sigma(R_{\hi})=1~\Msol~\rm{pc}^{-2}$ (shown in Figure \ref{fig:RC_and_SD} as the vertical red dotted line).  Following \citetalias{Deg2024}, we define the structural velocity, $V_{\hi}$, as the model velocity at $R_{\hi}$ (that is $V_{\hi}=V_{m}(R_{\hi})$).  This is shown as the horizontal red dotted line in the upper panel of Figure \ref{fig:RC_and_SD}. 
 We calculate these quantities for each of the bootstrapped fits and use these distributions to obtain $R_{\hi}=1.8\pm0.2~\rm{kpc}$ and $V_{\hi}=16\pm3~\kms$.  One technical point is that the $R_{\hi}$ listed in Table \ref{Fluffy_Table} has been beam corrected by subtracting off the beam size from the diameter in quadrature, 
 while the vertical line in Figure \ref{fig:RC_and_SD} has not as the entire profile has not been beam corrected.  The difference between the uncorrected and corrected sizes is $\sim0.1~\rm{kpc}$.


In addition to our best fitting model, Figure \ref{fig:RC_and_SD} shows the rotation curve and surface density profile for the \citetalias{Deg2024} model corrected to the TRGB distance (green) and then corrected to our measured inclination (blue).  Even with these corrections, the \citetalias{Deg2024} model has a slightly higher rotation curve and surface density profile.  This remaining difference is due to the differing center points used in the analysis.  Despite these differences, both models fit the data equally well (see Appendix \ref{app:ChannelMaps}).  However, our model has the same center as the stellar distribution, leading to its usage for further DM exploration.




\subsection{Stellar Models}\label{subsec:stellar_models}


The DM modelling of Sec. \ref{sec:DMcontent} also requires the stellar surface density profile.  We obtained this via an application of \textsc{AutoProf} to high-resolution $g,r,z$ band cutouts from the Dark Energy Sky Instrument Legacy Imaging Survey Date Release-10 \citep[hereafter DESI;][]{DESI, Dey2019}.  In order to generate a self-consistent model, the inclination and position angle are fixed to our best fitting parameters from the \hi\ kinematic analysis (Sec. \ref{ssec:KinModel}). The \textsc{AutoProf} surface brightness profiles are converted to a stellar surface density profile following the same methodology of \citet{Arora2021,Arora2023}, where a number of different mass-to-light ratios are applied based on the observed colors \citep{Courteau2014,Roediger2015,Zhang2017,Benito2019}. 
The use of multiple colors yields better constraints on mass-to-light ratios.
Ultimately, this approach yields 30 stellar surface density profiles and mass estimates that are averaged to provide the final surface density profile and stellar mass measurement of $\log_{10}(M_*/\Msol)=6.8\pm0.2$ used throughout this study.  The error in both quantities is the standard deviation of the 30 mass measurements and profiles.

The result of this approach is that we have built the most reliable kinematic model of KK 176 to date.  The model has, by construction, a consistent geometry for both the stellar and gaseous components.  And, as shown in Appendix \ref{app:ChannelMaps}, the fit to the data is excellent.  Additionally, given the quality of the data and the resolution of the observations, this is one of the most robust models of a diffuse dwarf available to date.

\section{Dark Matter Content} \label{sec:DMcontent}

There are two distinct questions about DM in diffuse dwarfs; do they have significant amounts of DM, and is their DM content consistent with expectations from scaling relations.  These questions can be explored via mass decomposition (Sec.~\ref{ssec:massmodel}) and the bTFR (Sec.~\ref{ssec:bTFR}). 

\subsection{Mass Decomposition}\label{ssec:massmodel}

In general, the DM content of a galaxy can be determined by comparing the total circular speed, $V_{\rm{circ}}$, to the contributions of the baryonic components:
\begin{equation}
    V_{\rm{circ}}^2 = V_{\rm{DM}}^2 + V_{\rm{S}}^2 + V_{\rm{g}}^2,
    \label{eq:massmodel}
\end{equation}
where $V_{\rm{DM}}$ is the contribution from the DM, $V_{\rm{S}}$ is the contribution from the stars, and $V_{\rm{g}}$ is the contribution from the gas. 
In low mass galaxies like KK 176, the relatively large velocity dispersion relative to the rotation velocity leads to asymmetric drift $V_{\rm{A}}$ which contributes pressure support.  The total circular speed due to gravity is then
\begin{equation}
    V_{\rm{circ}}^2 = V_{\rm{A}}^2+V_{\rm{m}}^2~.
\end{equation}
As shown in \citet{Iorio_2016}, $V_{A}$ can be estimated using
\begin{equation}\label{Eq:AsymmetricDrift}
    V_{\rm{A}}^2(R) = -\frac{R}{\rho}\frac{\partial\rho V_{\rm{disp}}^2}{\partial R} = -RV_{\rm{disp}}^2 \frac{\partial \ln(\rho V_{\rm{disp}}^2)}{\partial R}~, 
\end{equation}
where R is radius, $\rho=\Sigma(R)h(Z)$ is the volumetric density of the gas ($\Sigma(R)$ is the surface density and $h(Z)$ is the vertical density), and $V_{\rm{disp}}=8~\kms$ is the velocity dispersion adopted in the kinematic models (see Sec.~\ref{sec:modelling}). Separating the volumetric density and assuming a constant velocity dispersion means that only the derivative with respect to the surface density is necessary in Eq. \ref{Eq:AsymmetricDrift}.



We determined the stellar and gaseous rotation curve contributions using the \galpynamics\footnote{\url{https://gitlab.com/iogiul/galpynamics}} software package \citep{Iorio2018}.  This package fits a variety of different parametric profiles (i.e. exponential, Gaussian) to the measured surface density profiles and uses these to calculate the rotation velocity contributions from each component. For this work, we adopted the razor-thin exponential disk profiles for both the gaseous and stellar components, 
as this will maximize their contribution to the rotation curve, thereby minimizing the DM contribution. 

\begin{figure}
    \centering
    \includegraphics[width=1\linewidth]{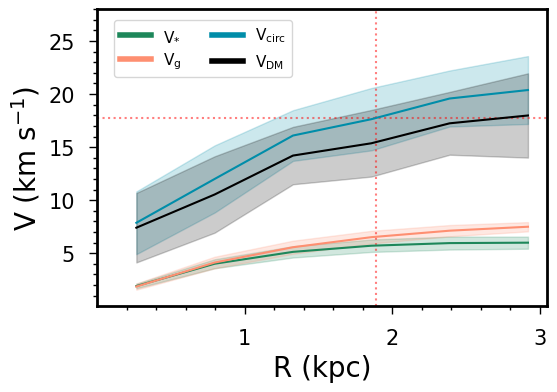}
    \caption{Mass decomposition of KK 176, showing the contributions to the circular velocity $V_{\rm{circ}}$ (blue curve) from different components (see Sec.~\ref{ssec:massmodel}). The contributions from the stars and the gas are shown in green and orange, respectively. The black curve shows the implied DM contribution (see Eq.~\ref{eq:massmodel}). The shaded regions surrounding each curve show standard deviations of the model components, estimated from bootstrap resampling (see Sec.~\ref{ssec:KinModel}).  The vertical and horizontal dotted red lines show $R_{\hi}$ and $V_{\rm{circ}}(R_{\hi})$ respectively.}
    \label{fig:MassModel}
\end{figure}

Figure \ref{fig:MassModel} shows the mass decomposition for KK 176. The uncertainties are calculated by performing a mass decomposition of each bootstrap sample and examining the distributions of each component.  From KK 176's mass decomposition, it can be seen that the \hi\ and the stellar component have similar contributions to the potential. 
The greater extent of the \hi\ mass in Table \ref{Fluffy_Table} is due to the greater extent of the \hi\ disk (see Figure \ref{fig:overlay} for a comparison). Nonetheless, both components are subdominant relative to the DM at all radii. The uncertainties also show that there is no version of a model that would have little to no DM. 


\subsection{The bTFR} \label{ssec:bTFR}

While KK 176 is dominated by DM at all radii, it is possible that it is still DM deficient relative to expectations. The bTFR is ultimately the relation between the baryonic mass and total mass of a galaxy.  Falling off the bTFR can imply that a system is either DM deficient or has too much DM (depending on where it falls).

If gas rich diffuse dwarfs fall on the bTFR, this would help extend the bTFR to lower mass regimes and help determine which types of low surface brightness objects fall on the bTFR.

To find KK 176's location on the bTFR, the total baryonic mass and a corresponding representative velocity must be calculated.  For kinematically modelled galaxies, the representative velocity can be derived from a few different parts of the rotation curve: the velocity of the `flat' portion of the rotation curve \citep{Lelli2019}, the outermost rotation point \citep{Ponomareva2021}, or the velocity at $R_{\hi}$ \citepalias{Deg2024}. For more poorly resolved galaxies, the rotation can be derived from spectra or from the position-velocity diagram \citep{McQuinn2022}.  We have adopted a similar approach to \citetalias{Deg2024}, where the velocity of KK 176 is its total circular speed at $R_{\hi}$. By considering the asymmetric drift and using the bootstrapped uncertainties, this gives a speed of $V_{\rm{circ}}(R_{\hi})=18\pm3~\kms$.

\begin{figure*}
    \centering
    \includegraphics[width=1\linewidth]{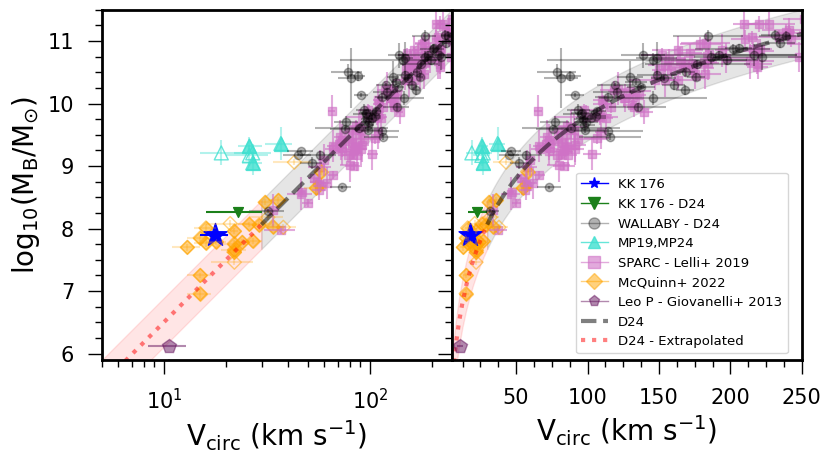}
    \caption{KK 176's position on the bTFR, along with other gas-rich galaxy samples from the literature. Both panels show the same data and relation, using either a logarthmic (panel A) or linear (panel B) stretch.  Our measurements of KK 176 are shown as a blue star. The green downward triangle shows the previous estimate from the pipeline-generated model in \citetalias{Deg2024} with a cosmic flow distance estimate instead of the TRGB distance $d$ adopted here. The SPARC data points from \citet{Lelli2019}, the WALLABY points from \citetalias{Deg2024}, and the UDGs from \citetalias{ManceraPina2019} (and the updated measurement for AGC 114905 from \citealt{Mancera_Pina_2024}) are all constructed using kinematic models.  By contrast the low mass \citet{McQuinn2022} measurements are from position-velocity (PV) diagram modelling (including an asymmetric drift correction).  Similarly the rotation velocity for Leo P is also determined from the PV diagram (via a slightly different technique), but does not include an asymmetric drift correction \citep{Giovanelli2013}. The SPARC and WALLABY points only show galaxies with $i\ge40^{\circ}$ (in accordance with (Deg+ 2025)).  Due to their lower sample sizes, we show the full \citetalias{ManceraPina2019} and \citet{McQuinn2022} samples, but their galaxies with $i<40^{\circ}$ are shown as open symbols.  We show the  bTFR scaling relation fit and associated scatter from \citetalias{Deg2024} as the dashed grey line and shaded region, and its extrapolation to the low mass regime as the dotted red line and shaded region.  Note that this fit is based solely on the grey WALLABY data points. KK 176, and the majority of the other systems at the low-mass end of the bTFR, are consistent with the \citetalias{Deg2024} estimate within its scatter. 
    }
    \label{fig:bTFR}
\end{figure*}

Figure \ref{fig:bTFR} shows the location of KK 176 on the bTFR compared to WALLABY models from \citetalias{Deg2024}, SPARC models from \citet{Lelli2019}, the gas-rich UDGs of \citetalias{ManceraPina2019}, 
the low mass dwarfs of \citet{McQuinn2022}, and the nearby, low mass dwarf Leo P \citep{Giovanelli2013}.  For AGC 114905 we use the updated circular speed from \citet{Mancera_Pina_2024}.  The left hand panel shows it in the usual log-log space, while the right-hand panel uses a linear scale for the velocities.  The different projections are intended to highlight that discrepancies of only a few $\kms$ at very low masses can lead to large differences from extrapolations of the bTFR in log-space projections.  This is particularly relevant in the low mass end, where different rotation definitions (i.e. $V_{\hi}$ compared to the flat part of the rotation curve compared to PV-diagrams) can cause significant, systematic changes to the bTFR \citep{Lelli2019}.  Nonetheless, both our model for KK 176 (blue star), as well as the previous \citetalias{Deg2024} model (green triangle) are consistent with the \citetalias{Deg2024} bTFR.  
KK 176 is also consistent with the low mass dwarfs of \citet{McQuinn2022} as well as the continued extrapolation down to Leo P \citep{Giovanelli2013}.
It is worth noting that the rotation velocities for the \cite{McQuinn2022} sample and Leo P were determined from PV diagram techniques. The \citet{McQuinn2022} sample includes an asymmetric drift correction, however it is not the same as the \citet{Iorio_2016} correction.
To ensure that the kinematic models are reliable, only WALLABY and SPARC points with $i\ge 40^{\circ}$ are shown in Figure \ref{fig:bTFR} (Deg et al., submitted).  The solid symbols for the \citet{McQuinn2022} and \citetalias{ManceraPina2019} samples also have $i\ge40^{\circ}$, while the open symbols have lower inclinations.



The UDGs of \citetalias{ManceraPina2019} are strong outliers from the bTFR and pose a significant challenge that remains to be solved. They are distinctly different from even the small outliers from \citet{McQuinn2022}, which, when examined in the right hand panel of Figure \ref{fig:bTFR}, lie within a few $\kms$ of the \citet{Deg2024} bTFR relation. The challenge of the \citetalias{ManceraPina2019} UDGs is made even more significant by the location of KK 176 on the bTFR. Works like \citet{Motiwala2025} suggest that gas rich diffuse dwarfs and gas rich UDGs form a continuum.  However, the gas-rich dwarf KK 176 lies on the bTFR while the gas rich UDGs of \citetalias{ManceraPina2019} do not. Determining whether KK 176 and the \citetalias{ManceraPina2019} UDGs are from distinctly different populations is beyond the scope of this work.  What is clear is that this well-resolved, isolated, nearby diffuse dwarf has a DM content that is fully consistent with other low mass dwarfs.


\section{Conclusion} \label{sec:conclusion}

KK 176 is an ideal labratory for the study of diffuse dwarfs owing to its proximity, location in a void, and well resolved \hi\ observations.  Using WALLABY observations we have constructed a strongly reliable kinematic model of a diffuse dwarf (see Sec. \ref{sec:modelling}, Appendix \ref{app:ChannelMaps}, and Figure \ref{fig:RC_and_SD}).  The model itself is anchored to the WISE center and has a consistent stellar and gaseous geometry.  We incorporated a `flipping bootstrap' approach to measure robust uncertainties for all model parameters (see Appendix \ref{apdix:bootstraps}).  This approach enables a straightforward method of calculating the uncertainties in all derived quantities by simply examining the distribution of bootstrapped fits. 

We utilized this kinematic model to do a mass decomposition for KK 176.  Accounting for the asymmetric drift due to the velocity dispersion, this model revealed that the DM component of KK 176 is dominant at all radii (see Sec. \ref{sec:DMcontent} and Figure \ref{fig:MassModel}).  

KK 176 is also consistent with the extrapolation of the \citetalias{Deg2024} bTFR and the \citet{McQuinn2022} low mass dwarfs (see Sec. \ref{sec:DMcontent} and Figure \ref{fig:bTFR}).  In other words, KK 176 has a rotation consistent with a `normal' 
amount of DM for a galaxy with its baryonic mass.  However, it remains possible that KK 176 may be an outlier in terms of the DM concentration.  This work has focused exclusively on measuring the DM content via a mass decomposition, but we have not fit a functional form to the DM profile.  Future observations with greater depth and resolution will be able to probe the inner regions of the galaxy and explore key questions like whether KK 176 has a core.  However, such an analysis is not within the scope of this work.

The fact that KK 176 has
a model with an improved distance and a center consistent with its stellar component, is isolated, is DM dominated, and lies on the bTFR heightens the tensions posed by the \citetalias{ManceraPina2019} UDGs.  It is possible that the \citetalias{ManceraPina2019} UDGs and KK 176 are unique classes of objects.  But such a claim is challenged by suggestions that diffuse dwarfs and UDGs form a continuum  \citep{Jones2023,Motiwala2025,Wright2025}.  Alternately, either KK 176 or the \citetalias{ManceraPina2019} UDGs could have been reconfigured by some sort of interaction or accretion event to move them either towards or away from the bTFR.  However, the fact that KK 176 lies in a void and is extremely metal poor suggests that it has not been reconfigured in recent time.  At this time, we cannot comment on the other possibility;  namely whether the \citetalias{ManceraPina2019} UDGs have undergone some recent reconfiguration that would move them away from the bTFR.  Fully resolving these tensions and understanding the connection between diffuse dwarfs, UDGs, DM, and the bTFR will require significantly larger sample sizes.

Such larger sample sizes will soon become available.  The wide area of WALLABY is already finding many low surface brightness galaxies, UDGs, and diffuse dwarfs \citep{OBeirne2025a}.  And this has been accomplished with only 1\% of the full survey coverage.  We expect to find many more of these objects in the future.  Some small fraction of these will be resolved as well as KK 176, enabling a statistical study of their DM properties. 
Follow-up surveys on instruments such as MeerKAT and the SKA will provide even greater sensitivity and resolution, allowing for studies of the low column density gas about KK 176 and other diffuse dwarfs and UDGs.  These will be critical for studying gas accretion, diffuse dwarf/UDG formation, and probing both their inner (with high resolution) and extended (with low column density) DM structures.

KK 176 is a robustly modelled, DM dominated, consistent with other low mass dwarfs, nearby diffuse dwarf which lies on the bTFR.  It is an anchor for future studies of diffuse dwarfs and UDGs.  It is the ideal labratory for future studies of DM structure, galaxy formation, and gas accretion/feedback.  Given the nature of WALLABY (coupled with deep optical observations like those provided by the Rubin Observatory), we expect to find many more KK 176's in the near future.

\section*{Acknowledgements}

We would like to thank A. Boselli, B. Holwerda, I. Karachentsev, T. Westmeier for useful comments and discussion.

PEMP is funded by the Dutch Research Council (NWO) through the Veni grant VI.Veni.222.364.

This scientific work uses data obtained from Inyarrimanha Ilgari Bundara / the Murchison Radio-astronomy Observatory. We acknowledge the Wajarri Yamaji People as the Traditional Owners and native title holders of the Observatory site. CSIRO’s ASKAP radio telescope is part of the Australia Telescope National Facility (https://ror.org/05qajvd42). Operation of ASKAP is funded by the Australian Government with support from the National Collaborative Research Infrastructure Strategy. ASKAP uses the resources of the Pawsey Supercomputing Research Centre. Establishment of ASKAP, Inyarrimanha Ilgari Bundara, the CSIRO Murchison Radio-astronomy Observatory and the Pawsey Supercomputing Research Centre are initiatives of the Australian Government, with support from the Government of Western Australia and the Science and Industry Endowment Fund.

This research was undertaken thanks in part to funding from the Canada First Research Excellence Fund through the Arthur B. McDonald Canadian Astroparticle Physics Research Institute.

This work was facilitated by the Australian SKA Regional Centre (AusSRC), Australia’s portion of the international SKA Regional Centre Network (SRCNet), funded by the Australian Government through the Department of Industry, Science, and Resources (DISR; grant SKARC000001). AusSRC is an equal collaboration between CSIRO – Australia’s national science agency, Curtin University, the Pawsey Supercomputing Research Centre, and the University of Western Australia.

%



\bibliography{fluffy}{}

\begin{thebibliography}{}
\expandafter\ifx\csname natexlab\endcsname\relax\def\natexlab#1{#1}\fi
\providecommand{\url}[1]{\href{#1}{#1}}
\providecommand{\dodoi}[1]{doi:~\href{http://doi.org/#1}{\nolinkurl{#1}}}
\providecommand{\doeprint}[1]{\href{http://ascl.net/#1}{\nolinkurl{http://ascl.net/#1}}}
\providecommand{\doarXiv}[1]{\href{https://arxiv.org/abs/#1}{\nolinkurl{https://arxiv.org/abs/#1}}}

\bibitem[{{Arnett}(1999)}]{Arnett1999}
{Arnett}, D. 1999, \apss, 265, 29, \dodoi{10.1023/A:1002131915162}

\bibitem[{{Arora} {et~al.}(2023){Arora}, {Courteau}, {Stone}, \& {Macci{\`o}}}]{Arora2023}
{Arora}, N., {Courteau}, S., {Stone}, C., \& {Macci{\`o}}, A.~V. 2023, \mnras, 522, 1208, \dodoi{10.1093/mnras/stad1023}

\bibitem[{{Arora} {et~al.}(2021){Arora}, {Stone}, {Courteau}, \& {Jarrett}}]{Arora2021}
{Arora}, N., {Stone}, C., {Courteau}, S., \& {Jarrett}, T.~H. 2021, \mnras, 505, 3135, \dodoi{10.1093/mnras/stab1430}

\bibitem[{Banik {et~al.}(2022)Banik, Nagesh, Haghi, Kroupa, \& Zhao}]{Banik_2022}
Banik, I., Nagesh, S.~T., Haghi, H., Kroupa, P., \& Zhao, H. 2022, Monthly Notices of the Royal Astronomical Society, 513, 3541–3548, \dodoi{10.1093/mnras/stac1073}

\bibitem[{{Buzzo} {et~al.}(2025){Buzzo}, {Forbes}, {Jarrett}, {Marleau}, {Duc}, {Brodie}, {Romanowsky}, {Ferr{\'e}-Mateu}, {Hilker}, {Gannon}, {Pfeffer}, \& {Haacke}}]{Buzzo2025}
{Buzzo}, M.~L., {Forbes}, D.~A., {Jarrett}, T.~H., {et~al.} 2025, \mnras, 536, 2536, \dodoi{10.1093/mnras/stae2700}

\bibitem[{{Comrie} {et~al.}(2021){Comrie}, {Wang}, {Hsu}, {Moraghan}, {Harris}, {Pang}, {Pi{\'n}ska}, {Chiang}, {Chang}, {Hwang}, {Jan}, {Lin}, \& {Simmonds}}]{CARTA}
{Comrie}, A., {Wang}, K.-S., {Hsu}, S.-C., {et~al.} 2021, {CARTA: The Cube Analysis and Rendering Tool for Astronomy}, 2.0.0,  Zenodo, \dodoi{10.5281/zenodo.3377984}

\bibitem[{{Courteau} {et~al.}(2014){Courteau}, {Cappellari}, {de Jong}, {Dutton}, {Emsellem}, {Hoekstra}, {Koopmans}, {Mamon}, {Maraston}, {Treu}, \& {Widrow}}]{Courteau2014}
{Courteau}, S., {Cappellari}, M., {de Jong}, R.~S., {et~al.} 2014, Reviews of Modern Physics, 86, 47, \dodoi{10.1103/RevModPhys.86.47}

\bibitem[{Deg {et~al.}(2022)Deg, Spekkens, Westmeier, Reynolds, Venkataraman, Goliath, Shen, Halloran, Bosma, Catinella, de~Blok, Dénes, DiTeodoro, Elagali, For, Howlett, Józsa, Kamphuis, Kleiner, Koribalski, Lee-Waddell, Lelli, Lin, Murugeshan, Oh, Rhee, Scott, Staveley-Smith, van~der Hulst, Verdes-Montenegro, Wang, \& Wong}]{Deg_2022}
Deg, N., Spekkens, K., Westmeier, T., {et~al.} 2022, Publications of the Astronomical Society of Australia, 39, \dodoi{10.1017/pasa.2022.43}

\bibitem[{{Deg} {et~al.}(2024){Deg}, {Arora}, {Spekkens}, {Halloran}, {Catinella}, {Jones}, {Courtois}, {Glazebrook}, {Bosma}, {Cortese}, {D{\'e}nes}, {Elagali}, {For}, {Kamphuis}, {Koribalski}, {Lee-Waddell}, {Mancera Pi{\~n}a}, {Mould}, {Rhee}, {Shao}, {Staveley-Smith}, {Wang}, {Westmeier}, \& {Wong}}]{Deg2024}
{Deg}, N., {Arora}, N., {Spekkens}, K., {et~al.} 2024, \apj, 976, 159, \dodoi{10.3847/1538-4357/ad84ba}

\bibitem[{{DESI Collaboration} {et~al.}(2016){DESI Collaboration}, {Aghamousa}, {Aguilar}, {Ahlen}, {Alam}, {Allen}, {Allende Prieto}, {Annis}, {Bailey}, {Balland}, {Ballester}, {Baltay}, {Beaufore}, {Bebek}, {Beers}, {Bell}, {Bernal}, {Besuner}, {Beutler}, {Blake}, {Bleuler}, {Blomqvist}, {Blum}, {Bolton}, {Briceno}, {Brooks}, {Brownstein}, {Buckley-Geer}, {Burden}, {Burtin}, {Busca}, {Cahn}, {Cai}, {Cardiel-Sas}, {Carlberg}, {Carton}, {Casas}, {Castander}, {Cervantes-Cota}, {Claybaugh}, {Close}, {Coker}, {Cole}, {Comparat}, {Cooper}, {Cousinou}, {Crocce}, {Cuby}, {Cunningham}, {Davis}, {Dawson}, {de la Macorra}, {De Vicente}, {Delubac}, {Derwent}, {Dey}, {Dhungana}, {Ding}, {Doel}, {Duan}, {Ealet}, {Edelstein}, {Eftekharzadeh}, {Eisenstein}, {Elliott}, {Escoffier}, {Evatt}, {Fagrelius}, {Fan}, {Fanning}, {Farahi}, {Farihi}, {Favole}, {Feng}, {Fernandez}, {Findlay}, {Finkbeiner}, {Fitzpatrick}, {Flaugher}, {Flender}, {Font-Ribera}, {Forero-Romero}, {Fosalba}, {Frenk}, {Fumagalli}, {Gaensicke}, {Gallo},
  {Garcia-Bellido}, {Gaztanaga}, {Pietro Gentile Fusillo}, {Gerard}, {Gershkovich}, {Giannantonio}, {Gillet}, {Gonzalez-de-Rivera}, {Gonzalez-Perez}, {Gott}, {Graur}, {Gutierrez}, {Guy}, {Habib}, {Heetderks}, {Heetderks}, {Heitmann}, {Hellwing}, {Herrera}, {Ho}, {Holland}, {Honscheid}, {Huff}, {Hutchinson}, {Huterer}, {Hwang}, {Illa Laguna}, {Ishikawa}, {Jacobs}, {Jeffrey}, {Jelinsky}, {Jennings}, {Jiang}, {Jimenez}, {Johnson}, {Joyce}, {Jullo}, {Juneau}, {Kama}, {Karcher}, {Karkar}, {Kehoe}, {Kennamer}, {Kent}, {Kilbinger}, {Kim}, {Kirkby}, {Kisner}, {Kitanidis}, {Kneib}, {Koposov}, {Kovacs}, {Koyama}, {Kremin}, {Kron}, {Kronig}, {Kueter-Young}, {Lacey}, {Lafever}, {Lahav}, {Lambert}, {Lampton}, {Landriau}, {Lang}, {Lauer}, {Le Goff}, {Le Guillou}, {Le Van Suu}, {Lee}, {Lee}, {Leitner}, {Lesser}, {Levi}, {L'Huillier}, {Li}, {Liang}, {Lin}, {Linder}, {Loebman}, {Luki{\'c}}, {Ma}, {MacCrann}, {Magneville}, {Makarem}, {Manera}, {Manser}, {Marshall}, {Martini}, {Massey}, {Matheson}, {McCauley}, {McDonald},
  {McGreer}, {Meisner}, {Metcalfe}, {Miller}, {Miquel}, {Moustakas}, {Myers}, {Naik}, {Newman}, {Nichol}, {Nicola}, {Nicolati da Costa}, {Nie}, {Niz}, {Norberg}, {Nord}, {Norman}, {Nugent}, {O'Brien}, {Oh}, {Olsen}, {Padilla}, {Padmanabhan}, {Padmanabhan}, {Palanque-Delabrouille}, {Palmese}, {Pappalardo}, {P{\^a}ris}, {Park}, {Patej}, {Peacock}, {Peiris}, {Peng}, {Percival}, {Perruchot}, {Pieri}, {Pogge}, {Pollack}, {Poppett}, {Prada}, {Prakash}, {Probst}, {Rabinowitz}, {Raichoor}, {Ree}, {Refregier}, {Regal}, {Reid}, {Reil}, {Rezaie}, {Rockosi}, {Roe}, {Ronayette}, {Roodman}, {Ross}, {Ross}, {Rossi}, {Rozo}, {Ruhlmann-Kleider}, {Rykoff}, {Sabiu}, {Samushia}, {Sanchez}, {Sanchez}, {Schlegel}, {Schneider}, {Schubnell}, {Secroun}, {Seljak}, {Seo}, {Serrano}, {Shafieloo}, {Shan}, {Sharples}, {Sholl}, {Shourt}, {Silber}, {Silva}, {Sirk}, {Slosar}, {Smith}, {Smoot}, {Som}, {Song}, {Sprayberry}, {Staten}, {Stefanik}, {Tarle}, {Sien Tie}, {Tinker}, {Tojeiro}, {Valdes}, {Valenzuela}, {Valluri}, {Vargas-Magana},
  {Verde}, {Walker}, {Wang}, {Wang}, {Weaver}, {Weaverdyck}, {Wechsler}, {Weinberg}, {White}, {Yang}, {Yeche}, {Zhang}, {Zhao}, {Zheng}, {Zhou}, {Zhou}, {Zhu}, {Zou}, \& {Zu}}]{DESI}
{DESI Collaboration}, {Aghamousa}, A., {Aguilar}, J., {et~al.} 2016, arXiv e-prints, arXiv:1611.00036, \dodoi{10.48550/arXiv.1611.00036}

\bibitem[{{Dey} {et~al.}(2019){Dey}, {Schlegel}, {Lang}, {Blum}, {Burleigh}, {Fan}, {Findlay}, {Finkbeiner}, {Herrera}, {Juneau}, {Landriau}, {Levi}, {McGreer}, {Meisner}, {Myers}, {Moustakas}, {Nugent}, {Patej}, {Schlafly}, {Walker}, {Valdes}, {Weaver}, {Y{\`e}che}, {Zou}, {Zhou}, {Abareshi}, {Abbott}, {Abolfathi}, {Aguilera}, {Alam}, {Allen}, {Alvarez}, {Annis}, {Ansarinejad}, {Aubert}, {Beechert}, {Bell}, {BenZvi}, {Beutler}, {Bielby}, {Bolton}, {Brice{\~n}o}, {Buckley-Geer}, {Butler}, {Calamida}, {Carlberg}, {Carter}, {Casas}, {Castander}, {Choi}, {Comparat}, {Cukanovaite}, {Delubac}, {DeVries}, {Dey}, {Dhungana}, {Dickinson}, {Ding}, {Donaldson}, {Duan}, {Duckworth}, {Eftekharzadeh}, {Eisenstein}, {Etourneau}, {Fagrelius}, {Farihi}, {Fitzpatrick}, {Font-Ribera}, {Fulmer}, {G{\"a}nsicke}, {Gaztanaga}, {George}, {Gerdes}, {Gontcho}, {Gorgoni}, {Green}, {Guy}, {Harmer}, {Hernandez}, {Honscheid}, {Huang}, {James}, {Jannuzi}, {Jiang}, {Joyce}, {Karcher}, {Karkar}, {Kehoe}, {Kneib}, {Kueter-Young}, {Lan},
  {Lauer}, {Le Guillou}, {Le Van Suu}, {Lee}, {Lesser}, {Perreault Levasseur}, {Li}, {Mann}, {Marshall}, {Mart{\'\i}nez-V{\'a}zquez}, {Martini}, {du Mas des Bourboux}, {McManus}, {Meier}, {M{\'e}nard}, {Metcalfe}, {Mu{\~n}oz-Guti{\'e}rrez}, {Najita}, {Napier}, {Narayan}, {Newman}, {Nie}, {Nord}, {Norman}, {Olsen}, {Paat}, {Palanque-Delabrouille}, {Peng}, {Poppett}, {Poremba}, {Prakash}, {Rabinowitz}, {Raichoor}, {Rezaie}, {Robertson}, {Roe}, {Ross}, {Ross}, {Rudnick}, {Safonova}, {Saha}, {S{\'a}nchez}, {Savary}, {Schweiker}, {Scott}, {Seo}, {Shan}, {Silva}, {Slepian}, {Soto}, {Sprayberry}, {Staten}, {Stillman}, {Stupak}, {Summers}, {Sien Tie}, {Tirado}, {Vargas-Maga{\~n}a}, {Vivas}, {Wechsler}, {Williams}, {Yang}, {Yang}, {Yapici}, {Zaritsky}, {Zenteno}, {Zhang}, {Zhang}, {Zhou}, \& {Zhou}}]{Dey2019}
{Dey}, A., {Schlegel}, D.~J., {Lang}, D., {et~al.} 2019, \aj, 157, 168, \dodoi{10.3847/1538-3881/ab089d}

\bibitem[{{Di Teodoro} \& {Fraternali}(2015)}]{diTeodoro15}
{Di Teodoro}, E.~M., \& {Fraternali}, F. 2015, \mnras, 451, 3021, \dodoi{10.1093/mnras/stv1213}

\bibitem[{{Du} {et~al.}(2024){Du}, {Du}, {Cheng}, {Zhu}, {Yu}, \& {Wu}}]{Du2024}
{Du}, L., {Du}, W., {Cheng}, C., {et~al.} 2024, \apj, 964, 85, \dodoi{10.3847/1538-4357/ad234f}

\bibitem[{{Garc{\'\i}a-Benito} {et~al.}(2019){Garc{\'\i}a-Benito}, {Gonz{\'a}lez Delgado}, {P{\'e}rez}, {Cid Fernandes}, {S{\'a}nchez}, \& {de Amorim}}]{Benito2019}
{Garc{\'\i}a-Benito}, R., {Gonz{\'a}lez Delgado}, R.~M., {P{\'e}rez}, E., {et~al.} 2019, \aap, 621, A120, \dodoi{10.1051/0004-6361/201833993}

\bibitem[{{Giovanelli} {et~al.}(2013){Giovanelli}, {Haynes}, {Adams}, {Cannon}, {Rhode}, {Salzer}, {Skillman}, {Bernstein-Cooper}, \& {McQuinn}}]{Giovanelli2013}
{Giovanelli}, R., {Haynes}, M.~P., {Adams}, E. A.~K., {et~al.} 2013, \aj, 146, 15, \dodoi{10.1088/0004-6256/146/1/15}

\bibitem[{{He} {et~al.}(2019){He}, {Wu}, {Du}, {Wicker}, {Zhao}, {Lei}, \& {Liu}}]{He2019}
{He}, M., {Wu}, H., {Du}, W., {et~al.} 2019, \apj, 880, 30, \dodoi{10.3847/1538-4357/ab2710}

\bibitem[{{Hotan} {et~al.}(2021){Hotan}, {Bunton}, {Chippendale}, {Whiting}, {Tuthill}, {Moss}, {McConnell}, {Amy}, {Huynh}, {Allison}, {Anderson}, {Bannister}, {Bastholm}, {Beresford}, {Bock}, {Bolton}, {Chapman}, {Chow}, {Collier}, {Cooray}, {Cornwell}, {Diamond}, {Edwards}, {Feain}, {Franzen}, {George}, {Gupta}, {Hampson}, {Harvey-Smith}, {Hayman}, {Heywood}, {Jacka}, {Jackson}, {Jackson}, {Jeganathan}, {Johnston}, {Kesteven}, {Kleiner}, {Koribalski}, {Lee-Waddell}, {Lenc}, {Lensson}, {Mackay}, {Mahony}, {McClure-Griffiths}, {McConigley}, {Mirtschin}, {Ng}, {Norris}, {Pearce}, {Phillips}, {Pilawa}, {Raja}, {Reynolds}, {Roberts}, {Roxby}, {Sadler}, {Shields}, {Schinckel}, {Serra}, {Shaw}, {Sweetnam}, {Troup}, {Tzioumis}, {Voronkov}, \& {Westmeier}}]{Hotan2021}
{Hotan}, A.~W., {Bunton}, J.~D., {Chippendale}, A.~P., {et~al.} 2021, \pasa, 38, e009, \dodoi{10.1017/pasa.2021.1}

\bibitem[{{Hu} {et~al.}(2023){Hu}, {Guo}, {Zheng}, {Yang}, {Tsai}, {Zhang}, \& {Zhang}}]{Hu2023}
{Hu}, H.-J., {Guo}, Q., {Zheng}, Z., {et~al.} 2023, \apjl, 947, L9, \dodoi{10.3847/2041-8213/acc7a4}

\bibitem[{Iorio(2018)}]{Iorio2018}
Iorio, G. 2018, PhD thesis, alma.
\newblock \url{https://amsdottorato.unibo.it/id/eprint/8449/}

\bibitem[{Iorio {et~al.}(2016)Iorio, Fraternali, Nipoti, Di~Teodoro, Read, \& Battaglia}]{Iorio_2016}
Iorio, G., Fraternali, F., Nipoti, C., {et~al.} 2016, Monthly Notices of the Royal Astronomical Society, stw3285, \dodoi{10.1093/mnras/stw3285}

\bibitem[{{Iorio} {et~al.}(2017){Iorio}, {Fraternali}, {Nipoti}, {Di Teodoro}, {Read}, \& {Battaglia}}]{Iorio2017}
{Iorio}, G., {Fraternali}, F., {Nipoti}, C., {et~al.} 2017, \mnras, 466, 4159, \dodoi{10.1093/mnras/stw3285}

\bibitem[{{Jones} {et~al.}(2023){Jones}, {Karunakaran}, {Bennet}, {Sand}, {Spekkens}, {Mutlu-Pakdil}, {Crnojevi{\'c}}, {Janowiecki}, {Leisman}, \& {Fielder}}]{Jones2023}
{Jones}, M.~G., {Karunakaran}, A., {Bennet}, P., {et~al.} 2023, \apjl, 942, L5, \dodoi{10.3847/2041-8213/acaaab}

\bibitem[{{Jones} {et~al.}(2025){Jones}, {Rey}, {Sand}, {Spekkens}, {Mutlu-Pakdil}, {Adams}, {Bennet}, {Crnojevic}, {Doliva-Dolinsky}, {Donnerstein}, {Fielder}, {Healy}, {Hunter}, {Karunakaran}, {Prabhu}, \& {Zaritsky}}]{Jones2025}
{Jones}, M.~G., {Rey}, M.~P., {Sand}, D.~J., {et~al.} 2025, arXiv e-prints, arXiv:2506.06424, \dodoi{10.48550/arXiv.2506.06424}

\bibitem[{{J{\'o}zsa} {et~al.}(2007){J{\'o}zsa}, {Kenn}, {Klein}, \& {Oosterloo}}]{Jorza2007}
{J{\'o}zsa}, G.~I.~G., {Kenn}, F., {Klein}, U., \& {Oosterloo}, T.~A. 2007, \aap, 468, 731, \dodoi{10.1051/0004-6361:20066164}

\bibitem[{{Kamphuis} {et~al.}(2015){Kamphuis}, {J{\'o}zsa}, {Oh}, {Spekkens}, {Urbancic}, {Serra}, {Koribalski}, \& {Dettmar}}]{Kamphuis2015}
{Kamphuis}, P., {J{\'o}zsa}, G.~I.~G., {Oh}, S. .~H., {et~al.} 2015, \mnras, 452, 3139, \dodoi{10.1093/mnras/stv1480}

\bibitem[{{Karachentsev} {et~al.}(2017){Karachentsev}, {Makarova}, {Tully}, {Rizzi}, {Karachentseva}, \& {Shaya}}]{Karachentsev17}
{Karachentsev}, I.~D., {Makarova}, L.~N., {Tully}, R.~B., {et~al.} 2017, \mnras, 469, L113, \dodoi{10.1093/mnrasl/slx061}

\bibitem[{{Karunakaran} {et~al.}(2022){Karunakaran}, {Spekkens}, {Carroll}, {Sand}, {Bennet}, {Crnojevi{\'c}}, {Jones}, \& {Mutlu-Pakd{\i}l}}]{Karunakaran2022}
{Karunakaran}, A., {Spekkens}, K., {Carroll}, R., {et~al.} 2022, \mnras, 516, 1741, \dodoi{10.1093/mnras/stac2329}

\bibitem[{{Koribalski} {et~al.}(2020){Koribalski}, {Staveley-Smith}, {Westmeier}, {Serra}, {Spekkens}, {Wong}, {Lee-Waddell}, {Lagos}, {Obreschkow}, {Ryan-Weber}, {Zwaan}, {Kilborn}, {Bekiaris}, {Bekki}, {Bigiel}, {Boselli}, {Bosma}, {Catinella}, {Chauhan}, {Cluver}, {Colless}, {Courtois}, {Crain}, {de Blok}, {D{\'e}nes}, {Duffy}, {Elagali}, {Fluke}, {For}, {Heald}, {Henning}, {Hess}, {Holwerda}, {Howlett}, {Jarrett}, {Jones}, {Jones}, {J{\'o}zsa}, {Jurek}, {J{\"u}tte}, {Kamphuis}, {Karachentsev}, {Kerp}, {Kleiner}, {Kraan-Korteweg}, {L{\'o}pez-S{\'a}nchez}, {Madrid}, {Meyer}, {Mould}, {Murugeshan}, {Norris}, {Oh}, {Oosterloo}, {Popping}, {Putman}, {Reynolds}, {Rhee}, {Robotham}, {Ryder}, {Schr{\"o}der}, {Shao}, {Stevens}, {Taylor}, {van{\^A} der Hulst}, {Verdes-Montenegro}, {Wakker}, {Wang}, {Whiting}, {Winkel}, \& {Wolf}}]{Koribalski2020}
{Koribalski}, B.~S., {Staveley-Smith}, L., {Westmeier}, T., {et~al.} 2020, \apss, 365, 118, \dodoi{10.1007/s10509-020-03831-4}

\bibitem[{{Kourkchi} {et~al.}(2020){Kourkchi}, {Courtois}, {Graziani}, {Hoffman}, {Pomar{\`e}de}, {Shaya}, \& {Tully}}]{Kourkchi2020}
{Kourkchi}, E., {Courtois}, H.~M., {Graziani}, R., {et~al.} 2020, \aj, 159, 67, \dodoi{10.3847/1538-3881/ab620e}

\bibitem[{{Kurapati} {et~al.}(2024){Kurapati}, {Pustilnik}, \& {Egorova}}]{Kurapati2024}
{Kurapati}, S., {Pustilnik}, S.~A., \& {Egorova}, E.~S. 2024, \mnras, 533, 1178, \dodoi{10.1093/mnras/stae1894}

\bibitem[{{Kuzio de Naray} {et~al.}(2006){Kuzio de Naray}, {McGaugh}, {de Blok}, \& {Bosma}}]{KuzioDeNaray2006}
{Kuzio de Naray}, R., {McGaugh}, S.~S., {de Blok}, W.~J.~G., \& {Bosma}, A. 2006, \apjs, 165, 461, \dodoi{10.1086/505345}

\bibitem[{{Kuzio de Naray} \& {Spekkens}(2011)}]{KuzioDeNaray2011}
{Kuzio de Naray}, R., \& {Spekkens}, K. 2011, \apjl, 741, L29, \dodoi{10.1088/2041-8205/741/2/L29}

\bibitem[{{Lelli} {et~al.}(2019){Lelli}, {McGaugh}, {Schombert}, {Desmond}, \& {Katz}}]{Lelli2019}
{Lelli}, F., {McGaugh}, S.~S., {Schombert}, J.~M., {Desmond}, H., \& {Katz}, H. 2019, \mnras, 484, 3267, \dodoi{10.1093/mnras/stz205}

\bibitem[{{Lim} {et~al.}(2020){Lim}, {C{\^o}t{\'e}}, {Peng}, {Ferrarese}, {Roediger}, {Durrell}, {Mihos}, {Wang}, {Gwyn}, {Cuillandre}, {Liu}, {S{\'a}nchez-Janssen}, {Toloba}, {Sales}, {Guhathakurta}, {Lan{\c{c}}on}, \& {Puzia}}]{Lim2020}
{Lim}, S., {C{\^o}t{\'e}}, P., {Peng}, E.~W., {et~al.} 2020, \apj, 899, 69, \dodoi{10.3847/1538-4357/aba433}

\bibitem[{{Mancera Pi{\~n}a} {et~al.}(2022){Mancera Pi{\~n}a}, {Fraternali}, {Oosterloo}, {Adams}, {Oman}, \& {Leisman}}]{mancera_pina_2022}
{Mancera Pi{\~n}a}, P.~E., {Fraternali}, F., {Oosterloo}, T., {et~al.} 2022, \mnras, 512, 3230, \dodoi{10.1093/mnras/stab3491}

\bibitem[{{Mancera Pi{\~n}a} {et~al.}(2025){Mancera Pi{\~n}a}, {Read}, {Kim}, {Marasco}, {Benavides}, {Glowacki}, {Pezzulli}, \& {Lagos}}]{mancerapina_2025}
{Mancera Pi{\~n}a}, P.~E., {Read}, J.~I., {Kim}, S., {et~al.} 2025, \aap, 699, A311, \dodoi{10.1051/0004-6361/202554381}

\bibitem[{{Mancera Pi{\~n}a} {et~al.}(2019){Mancera Pi{\~n}a}, {Fraternali}, {Adams}, {Marasco}, {Oosterloo}, {Oman}, {Leisman}, {di Teodoro}, {Posti}, {Battipaglia}, {Cannon}, {Gault}, {Haynes}, {Janowiecki}, {McAllan}, {Pagel}, {Reiter}, {Rhode}, {Salzer}, \& {Smith}}]{ManceraPina2019}
{Mancera Pi{\~n}a}, P.~E., {Fraternali}, F., {Adams}, E. A.~K., {et~al.} 2019, \apjl, 883, L33, \dodoi{10.3847/2041-8213/ab40c7}

\bibitem[{Mancera~Piña {et~al.}(2024)Mancera~Piña, Golini, Trujillo, \& Montes}]{Mancera_Pina_2024}
Mancera~Piña, P.~E., Golini, G., Trujillo, I., \& Montes, M. 2024, \aap, 689, A344, \dodoi{10.1051/0004-6361/202450230}

\bibitem[{{McQuinn} {et~al.}(2022){McQuinn}, {Adams}, {Cannon}, {Fuson}, {Skillman}, {Brooks}, {Rhode}, {Haynes}, {Inoue}, {Marine}, {Salzer}, \& {Talluri}}]{McQuinn2022}
{McQuinn}, K. B.~W., {Adams}, E. A.~K., {Cannon}, J.~M., {et~al.} 2022, \apj, 940, 8, \dodoi{10.3847/1538-4357/ac9285}

\bibitem[{{Motiwala} {et~al.}(2025){Motiwala}, {Karunakaran}, {Spekkens}, {Arora}, {Di Cintio}, {Wright}, {Zaritsky}, \& {Macci{\`o}}}]{Motiwala2025}
{Motiwala}, K., {Karunakaran}, A., {Spekkens}, K., {et~al.} 2025, arXiv e-prints, arXiv:2502.14971, \dodoi{10.48550/arXiv.2502.14971}

\bibitem[{{Murugeshan} {et~al.}(2024){Murugeshan}, {Deg}, {Westmeier}, {Shen}, {For}, {Spekkens}, {Wong}, {Staveley-Smith}, {Catinella}, {Lee-Waddell}, {D{\'e}nes}, {Rhee}, {Cortese}, {Goliath}, {Halloran}, {van der Hulst}, {Kamphuis}, {Koribalski}, {Kraan-Korteweg}, {Lelli}, {Venkataraman}, {Verdes-Montenegro}, \& {Yu}}]{Murugeshan2024}
{Murugeshan}, C., {Deg}, N., {Westmeier}, T., {et~al.} 2024, \pasa, 41, e088, \dodoi{10.1017/pasa.2024.91}

\bibitem[{{O'Beirne} {et~al.}(2025){O'Beirne}, {Staveley-Smith}, {Kilborn}, {Wong}, {Westmeier}, {Cluver}, {Bekki}, {Deg}, {D{\'e}nes}, {For}, {Lee-Waddell}, {Murugeshan}, {Oman}, {Rhee}, {Shen}, \& {Taylor}}]{OBeirne2025a}
{O'Beirne}, T., {Staveley-Smith}, L., {Kilborn}, V.~A., {et~al.} 2025, arXiv e-prints, arXiv:2505.04299, \dodoi{10.48550/arXiv.2505.04299}

\bibitem[{{Oh} {et~al.}(2015){Oh}, {Hunter}, {Brinks}, {Elmegreen}, {Schruba}, {Walter}, {Rupen}, {Young}, {Simpson}, {Johnson}, {Herrmann}, {Ficut-Vicas}, {Cigan}, {Heesen}, {Ashley}, \& {Zhang}}]{Oh2015}
{Oh}, S.-H., {Hunter}, D.~A., {Brinks}, E., {et~al.} 2015, \aj, 149, 180, \dodoi{10.1088/0004-6256/149/6/180}

\bibitem[{{Ponomareva} {et~al.}(2021){Ponomareva}, {Mulaudzi}, {Maddox}, {Frank}, {Jarvis}, {Di Teodoro}, {Glowacki}, {Kraan-Korteweg}, {Oosterloo}, {Adams}, {Pan}, {Prandoni}, {Rajohnson}, {Sinigaglia}, {Adams}, {Heywood}, {Bowler}, {Hatfield}, {Collier}, \& {Sekhar}}]{Ponomareva2021}
{Ponomareva}, A.~A., {Mulaudzi}, W., {Maddox}, N., {et~al.} 2021, \mnras, 508, 1195, \dodoi{10.1093/mnras/stab2654}

\bibitem[{{Pustilnik} {et~al.}(2021){Pustilnik}, {Egorova}, {Kniazev}, {Perepelitsyna}, {Tepliakova}, {Burenkov}, \& {Oparin}}]{Pustilnik2021}
{Pustilnik}, S.~A., {Egorova}, E.~S., {Kniazev}, A.~Y., {et~al.} 2021, \mnras, 507, 944, \dodoi{10.1093/mnras/stab2084}

\bibitem[{{Pustilnik} {et~al.}(2020){Pustilnik}, {Kniazev}, {Perepelitsyna}, \& {Egorova}}]{Pustilnik2020}
{Pustilnik}, S.~A., {Kniazev}, A.~Y., {Perepelitsyna}, Y.~A., \& {Egorova}, E.~S. 2020, \mnras, 493, 830, \dodoi{10.1093/mnras/staa215}

\bibitem[{{Read} {et~al.}(2016){Read}, {Iorio}, {Agertz}, \& {Fraternali}}]{Read2016}
{Read}, J.~I., {Iorio}, G., {Agertz}, O., \& {Fraternali}, F. 2016, \mnras, 462, 3628, \dodoi{10.1093/mnras/stw1876}

\bibitem[{{Roediger} \& {Courteau}(2015)}]{Roediger2015}
{Roediger}, J.~C., \& {Courteau}, S. 2015, \mnras, 452, 3209, \dodoi{10.1093/mnras/stv1499}

\bibitem[{{Sales} {et~al.}(2022){Sales}, {Wetzel}, \& {Fattahi}}]{Sales2022}
{Sales}, L.~V., {Wetzel}, A., \& {Fattahi}, A. 2022, Nature Astronomy, 6, 897, \dodoi{10.1038/s41550-022-01689-w}

\bibitem[{{Scott} {et~al.}(2021){Scott}, {Sengupta}, {Lagos}, {Chung}, \& {Wong}}]{Scott2021}
{Scott}, T.~C., {Sengupta}, C., {Lagos}, P., {Chung}, A., \& {Wong}, O.~I. 2021, \mnras, 503, 3953, \dodoi{10.1093/mnras/stab390}

\bibitem[{{Sengupta} {et~al.}(2019){Sengupta}, {Scott}, {Chung}, \& {Wong}}]{Sengupta2019}
{Sengupta}, C., {Scott}, T.~C., {Chung}, A., \& {Wong}, O.~I. 2019, \mnras, 488, 3222, \dodoi{10.1093/mnras/stz1884}

\bibitem[{{Serra} {et~al.}(2015){Serra}, {Westmeier}, {Giese}, {Jurek}, {Fl{\"o}er}, {Popping}, {Winkel}, {van der Hulst}, {Meyer}, {Koribalski}, {Staveley-Smith}, \& {Courtois}}]{Serra2015}
{Serra}, P., {Westmeier}, T., {Giese}, N., {et~al.} 2015, \mnras, 448, 1922, \dodoi{10.1093/mnras/stv079}

\bibitem[{{Spekkens} \& {Sellwood}(2007)}]{Spekkens2007}
{Spekkens}, K., \& {Sellwood}, J.~A. 2007, \apj, 664, 204, \dodoi{10.1086/518471}

\bibitem[{{Stone} {et~al.}(2021){Stone}, {Arora}, {Courteau}, \& {Cuillandre}}]{Stone2021}
{Stone}, C.~J., {Arora}, N., {Courteau}, S., \& {Cuillandre}, J.-C. 2021, \mnras, 508, 1870, \dodoi{10.1093/mnras/stab2709}

\bibitem[{{The GIMP Development Team}(2025)}]{GIMP}
{The GIMP Development Team}. 2025, GNU Image Manipulation Program (GIMP), Version 3.0.4. Community, Free Software (license GPLv3).
\newblock \url{https://gimp.org/}

\bibitem[{{van Dokkum} {et~al.}(2015){van Dokkum}, {Abraham}, {Merritt}, {Zhang}, {Geha}, \& {Conroy}}]{van_dokkum_2015}
{van Dokkum}, P.~G., {Abraham}, R., {Merritt}, A., {et~al.} 2015, \apjl, 798, L45, \dodoi{10.1088/2041-8205/798/2/L45}

\bibitem[{{Westmeier} {et~al.}(2021){Westmeier}, {Kitaeff}, {Pallot}, {Serra}, {van der Hulst}, {Jurek}, {Elagali}, {For}, {Kleiner}, {Koribalski}, {Lee-Waddell}, {Mould}, {Reynolds}, {Rhee}, \& {Staveley-Smith}}]{Westmeier2021}
{Westmeier}, T., {Kitaeff}, S., {Pallot}, D., {et~al.} 2021, \mnras, 506, 3962, \dodoi{10.1093/mnras/stab1881}

\bibitem[{{Wright} {et~al.}(2025){Wright}, {Brooks}, {Tremmel}, {Young}, {Munshi}, \& {Quinn}}]{Wright2025}
{Wright}, A.~C., {Brooks}, A.~M., {Tremmel}, M., {et~al.} 2025, arXiv e-prints, arXiv:2507.21231, \dodoi{10.48550/arXiv.2507.21231}

\bibitem[{{Wright} {et~al.}(2010){Wright}, {Eisenhardt}, {Mainzer}, {Ressler}, {Cutri}, {Jarrett}, {Kirkpatrick}, {Padgett}, {McMillan}, {Skrutskie}, {Stanford}, {Cohen}, {Walker}, {Mather}, {Leisawitz}, {Gautier}, {McLean}, {Benford}, {Lonsdale}, {Blain}, {Mendez}, {Irace}, {Duval}, {Liu}, {Royer}, {Heinrichsen}, {Howard}, {Shannon}, {Kendall}, {Walsh}, {Larsen}, {Cardon}, {Schick}, {Schwalm}, {Abid}, {Fabinsky}, {Naes}, \& {Tsai}}]{Wright2010}
{Wright}, E.~L., {Eisenhardt}, P. R.~M., {Mainzer}, A.~K., {et~al.} 2010, \aj, 140, 1868, \dodoi{10.1088/0004-6256/140/6/1868}

\bibitem[{{Zhang} {et~al.}(2017){Zhang}, {Puzia}, \& {Weisz}}]{Zhang2017}
{Zhang}, H.-X., {Puzia}, T.~H., \& {Weisz}, D.~R. 2017, \apjs, 233, 13, \dodoi{10.3847/1538-4365/aa937b}

\end{thebibliography}
\bibliographystyle{aasjournal}

\appendix

\section{Channel Maps and Model Residuals}\label{app:ChannelMaps}

\begin{figure*}[h]\label{apdix:channels}
    \centering
    \includegraphics[width=1\linewidth]{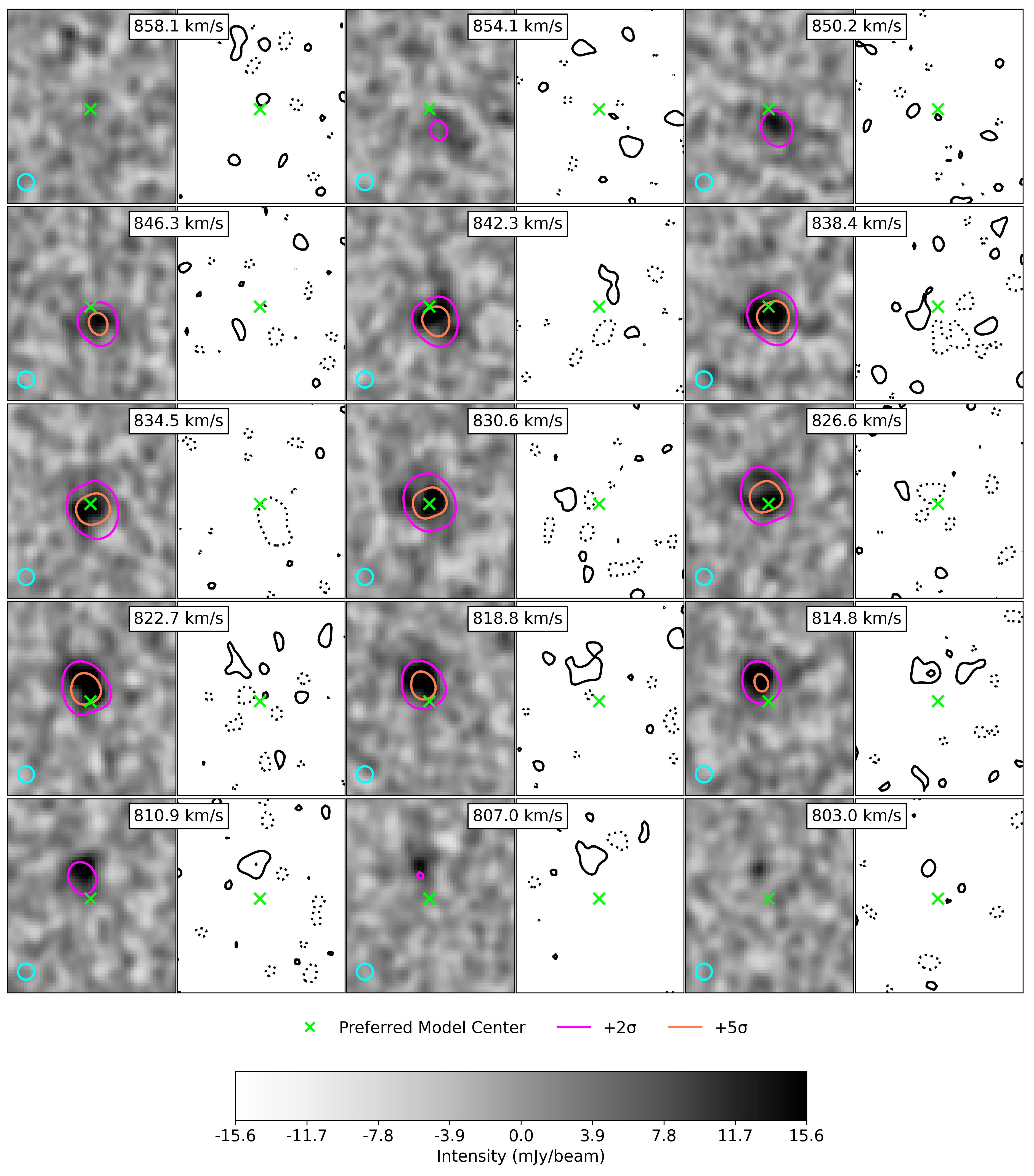}
    \caption{Pairs of \hi\ maps (left panels) and data -- model residuals for our preferred kinematic model (right panels) of Fluffy, labeled by each channel's heliocentric \hi\ recessional velocity. In all panels, the green cross shows our preferred model center measured from the WISE data (see Sec.~\ref{sec:modelling}). In the left panels, the cyan circle shows the WALLABY beam, and the magenta and orange lines show isodensity contours of our preferred model at $2\sigma = 5.2\,$mJy/beam and $5\sigma = 13\,$mJy/beam, respectively, where $\sigma$ is the RMS map noise. In the right panels, the solid and dashed lines show data -- model residuals at the $+2\sigma$ and $-2\sigma$ levels, respectively. } 
    \label{fig:enter-label}
\end{figure*}

To demonstrate the quality of our fit to the data, Figure \ref{fig:enter-label} compares the observed channel maps to our preferred model.
The left hand of each panel pair shows the data and model contour, while the right hand shows the data-model residual.
We also overplot the centre from our preferred model (green crosses). Our preferred model is a good representation of the emission in each channel, and its differences from the model in \citetalias{Deg2024} are relatively minor (see also Fig.~\ref{fig:RC_and_SD}). 


\section{Flipping Bootstrap Approach to Uncertainty Estimation}\label{apdix:bootstraps}

To estimate reliable kinematic model uncertainties, we have developed a `flipping' bootstrap approach. Although initially designed for use with the 3D-Kinematic Data aNalysis Algorithim for Surveys (3KIDNAS, Deg et al., in prep), we have adapted it for use with \barolo.  The `flipping' bootstrap approach provides statistically robust uncertainties for model parameters that can be easily propagated to model derived quantities by examining the distribution of those quantities calculated from all the bootstrapped fits.  The idea behind the flipping bootstrap is to generate some number of bootstrap realizations from the preferred model. From here, each bootstrap realization is re-fit with the same kinematic model as the data (in this case, with \barolo as described in Sec.~\ref{sec:modelling}), and the distributions of best-fitting parameters are used to compute the uncertainties on them. The challenge is to generate realistic bootstrap realizations, while also maintaining the coherence of any extended structures in the residuals. The `flipping' portion of the bootstrap is intended to accomplish this by flipping residuals in a channel about the axes of symmetry of an axisymmetric flat-disk model that are defined by the major and minor axes of the preferred model.

A bootstrap realization is generated by:
\begin{enumerate}
    \item Calculating a residual cube by subtracting the best fitting model from the data;
    \item Constructing a new residual cube from the data -- model residual cube.  For each channel group, residuals may be flipped across the major or minor axis of the preferred model.  When flipped over the minor axis, it is also necessary to flip across the systemic velocity to relocate residuals in the new realization in an equally (un-) likely location as in the data -- model residual cube.
    \item Add the best fiting model back into the new residual cube to generate a bootstrap realization.
\end{enumerate}
This process can generate up to $(N/M)!(N/2M)!$ unique bootstrap realizations of the data, where $N$ is the number of channels and $M$ is the number of grouped channels (in case one wants to preserve structures across multiple channels).  The one caveat is that in each cube, a particular residual feature located in a specific quadrant can only appear in one of four distinct locations in the bootstrap samples.  

Once fits for the $N$ bootstrap realizations of the preferred model to the data are in hand, the uncertainty $\Delta$ of a given model parameter is set to
\begin{equation}\label{Eq:BSUncertainty}
    \Delta^{2}=\rm{RMS(\mathbf{y})}^2+(x-\bar{\mathbf{y}})^2~,
\end{equation}
where $x$ is the best fitting value of that parameter, $\rm{RMS}(\mathbf{y})$ is the standard deviation of the distribution $\mathbf{y}$ of $N$ fitted values of that parameter from the bootstrap realizations, and $\bar{\mathbf{y}}$ is the mean of the distribution.  The second term accounts for any systematic offsets between the mean of the distribution of fitted values to the bootstrap realizations and the preferred model value of the data. Deg et al. (in prep) examine results from these bootstrap samples and found that these uncertainties correspond with the standard deviation of the distribution of preferred model values of the same underlying toy model with different noise realizations.

\begin{figure}[h]
    \centering
    \includegraphics[width=0.4\columnwidth]{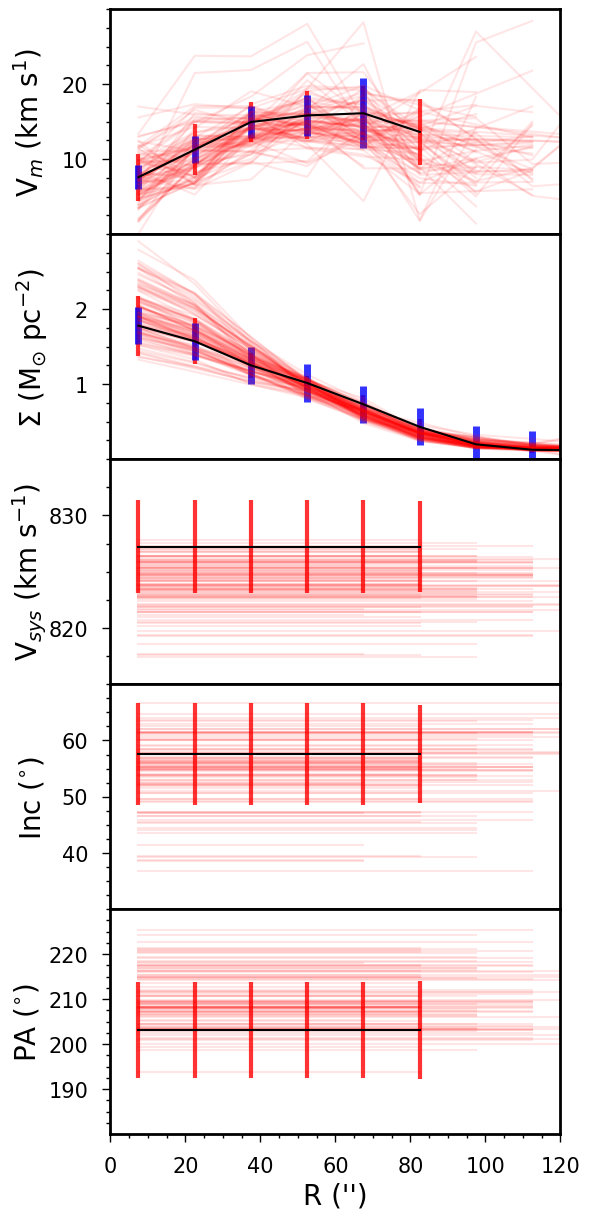}
    \caption{Our best fitting kinematic model for Fluffy from \barolo (black line, see Sec.~\ref{sec:modelling}) compared to the values returned from 100 bootstrapped realizations of that model (light red lines).  The blue error bars show the \barolo\ uncertainties for the rotation curve and surface density profiles, while the red errorbars show the bootstrapped uncertainties.}
    \label{fig:bootstrapdist}
\end{figure}

To illustrate the bootstrapping method for our kinematic models of Fluffy (see Sec.~\ref{sec:modelling}), Figure~\ref{fig:bootstrapdist} shows a comparison of the \barolo\ best fitting model to the distribution of 100 bootstrap fits.  It also shows the comparison of the bootstrapped uncertainties to those determined by the built in \barolo\ uncertainty estimate for the rotation curve and surface density profile.  There are a few key points raised by this figure. Firstly, some of the bootstrapped fits extend further than other fits.  This is primarily due to the $S/N$ of the observation coupled with the complexities of source finding.  Secondly, in the inclination and position angle panels, systematic offsets can be seen between the bootstrapped distribution and the initial \barolo\ fit.  This difference is precisely the reason for the second term in Eq. \ref{Eq:BSUncertainty} and highlights possible biases in the fitting.  Finally, and perhaps most importantly, the \barolo\ and bootstrapped uncertainties for the rotation curve and surface density profile are generally similar, except in the inner regions where the bootstrapped uncertainties are slightly larger.  Ultimately, Figure \ref{fig:bootstrapdist} shows that the flipping bootstrap does indeed produce reliable uncertainties for the full range of model parameters.  And, as noted above, this approach allows for straightforward propagation of uncertainties through to derived quantities.  For these reasons we prefer the flipping bootstrap uncertainties for the analysis of Fluffy.

\end{document}